\documentclass[aps,pra,floatfix,groupedaddress,titlepage,twocolumn,amsmath,amssymb,longbibliography]{revtex4-1}
 
\usepackage{graphicx}
\usepackage{bm}
\usepackage{braket}
\usepackage{times}
\expandafter\let\csname equation*\endcsname\relax
\expandafter\let\csname endequation*\endcsname\relax
\usepackage{amsmath}
\usepackage{amssymb}
\usepackage{amstext}  

\usepackage{color}
\definecolor{darkblue}{rgb}{0.2,0.2,0.8}
\usepackage[pdftex,
        colorlinks=true,
        linkcolor=darkblue,
        citecolor=darkblue,
        filecolor=black,
        urlcolor=darkblue,
        bookmarks=false,
        bookmarksopen=false,
        bookmarksopenlevel=3,
        plainpages=false,
        pdfpagelabels=true,
        verbose=true]{hyperref}
\usepackage[all]{hypcap}

\renewcommand{\:}{\,$$=$$\,}

\newcommand{\expect}[1]{\langle {#1} \rangle}
\newcommand{\Fig}[1]{Fig.~\ref{#1}}
\newcommand{\Ref}[1]{Ref.~\cite{#1}}

\begin{document}

\title{Twisted superfluid phase in the extended one-dimensional Bose--Hubbard model}

\newcommand{\ILP}{\affiliation{Institut f\"ur Laserphysik, Universit\"at Hamburg, Luruper Chaussee 149, 22761 Hamburg, Germany}}

\author{Dirk-S\"oren L\"uhmann}\ILP

\begin{abstract}
In one-dimensional systems a twisted superfluid phase is found which is induced by a spontaneous breaking of the time-reversal symmetry. Using the density-matrix renormalization group allows us to show that the excitation energy gap closes exponentially causing a quasi-degenerate ground state. The two degenerate ground states are connected by the time-reversal symmetry which manifests itself in an alternating complex phase of the long-range correlation function. The quantum phase transition to the twisted superfluid is driven by pair tunneling processes in an extended Bose--Hubbard model. The phase boundaries of several other phases are discussed including a supersolid, a pair superfluid, and a pair supersolid phase as well as a highly unconventional Mott insulator with a degenerate ground state and a staggered pair correlation function.
\end{abstract}   
 
\maketitle

One-dimensional bosonic systems are believed to be widely understood due to the unique theoretical insight by means of the density-matrix renormalization group (DMRG). This is even more remarkable as correlation effects in one-dimension are considerably stronger than in higher dimensions. It was shown that a Berezinskii--Kosterlitz--Thouless transition with an algebraically decaying long-range correlation function occurs between the superfluid and the Mott insulator phase \cite{Kuhner1998,Kuhner2000}. Additional nearest-neighbor processes, included in so-called extended Bose--Hubbard models \cite{Dutta2015}, lead to more complex phase diagrams including (charge) density waves with alternating integer occupations. Superfluidity combined with a density modulation indicating a supersolid behavior was not found for commensurate filling. However, for non-commensurate filling both a supersolid phase and a phase separation were predicted using quantum Monte-Carlo \cite{Batrouni2006} and DMRG methods \cite{Mishra2009}. Later a bosonic Haldane insulator was found that is characterized by non-local string correlations \cite{DallaTorre2006,Deng2011,Rossini2012}.
 
 Here, it is shown that an additional quantum phase can occur within the framework of extended Hubbard models. The phase transition is driven by correlated pair tunneling and is characterized by a spontaneous breaking of the time-reversal symmetry. In one dimension, this quantum phase can be identified by an alternating complex phase of the long-range correlation function.  For real Hamiltonians, Feynman's no-node theorem \cite{Feynman1972,Wu2009} states that the ground state of a bosonic system is positive real and therefore non-degenerate. However, the quasi-exact treatment of a large system with DMRG allows us to show that the excitation gap closes exponentially with an increasing system size. This leads to quasi-degenerate ground states for already relatively small systems. In the thermodynamic limit, this excitation gap becomes arbitrarily small. I show how the two lowest, real-valued DMRG states are connected to the complex-valued, time-reversal-broken ground states. Previously, this so-called twisted superfluid quantum phase was predicted only for honeycomb lattices loaded with single- and two-component bosons \cite{Soltan-Panahi2012,Jurgensen2015}. The mean-field treatment in \Ref{Jurgensen2015} intrinsically realizes the thermodynamic limit and is unable to access the energy gap, any system-length-dependent properties, and long-range correlation functions. For strong pair tunneling and nearest-neighbor interaction, the twisted superfluid phase becomes unstable against either a supersolid phase \cite{Sengupta2005,Batrouni2006,Mishra2009,Mazzarella2006,Wessel2007,Gan2007a,Gan2007b,Mathey2009,Maik2013,Jurgensen2015} or a pair superfluid  \cite{Kuklov2004,Mathey2009,Hu2009,Menotti2010,Sowinski2012,Jurgensen2015}. For small tunneling amplitudes, an unconventional type of Mott insulator appears located between the normal Mott insulator and the pair superfluid. It shows a degenerate ground state and a staggered pair correlation function. 

\begin{figure}[b!]
\includegraphics[width=\linewidth]{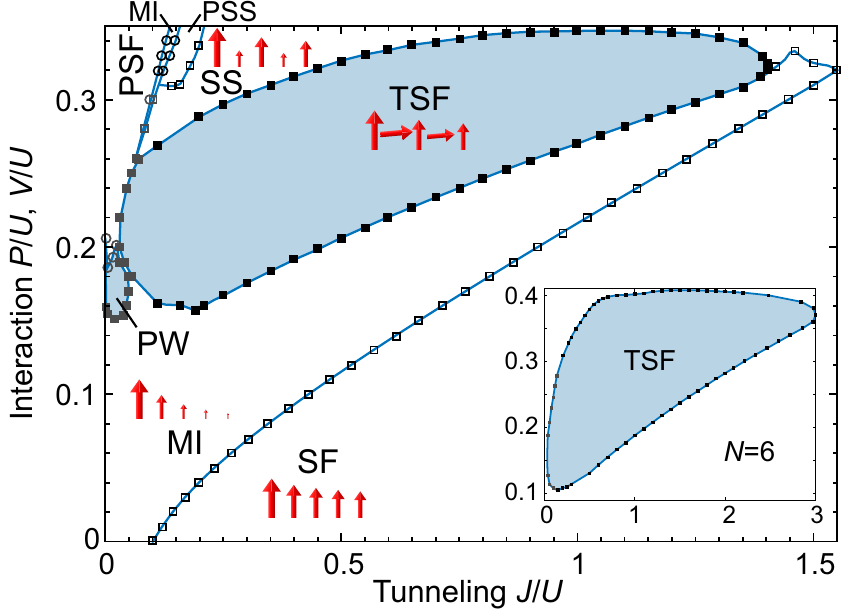}
\caption{DMRG phase diagram of the extended Hubbard model for a filling of $N=4$ particles per lattice site. The depicted phases are the superfluid (SF), twisted superfluid (TSF), Mott insulator (MI), supersolid (SS), pair superfluid (PSF), pair supersolid (PSS), and pair-wave Mott insulator (PW). The TSF and the PW phase are characterized by a degenerate ground state (shaded areas, where the filled markers represent an excitation gap smaller than $10^{-5}U$). The boundaries between MI and SF/SS are defined by a critical exponent $\nu_1=1/4$ of the long-range correlation function $g_1(j)$ (the open squares are obtained for $8$$\leq$$j$$\leq$$16$). The pair superfluid phase corresponds to $\nu_1<1/4$ and a critical exponent $\nu_p>1/4$ of the pair correlation function (open circles). The red arrows symbolize the decay of the correlation function $g_1(j)=\expect{\hat a_0^\dagger \hat a_j}$ at a distance $j\:0,...,4$ for different exponents $\nu_1$, where the vectors visualize amplitude and argument within the complex plane. The inset shows that the twisted superfluid phase extends substantially in parameter space for $N=6$ particles per site. Black (gray) markers correspond to 450 (256) states of the density matrix kept in each DMRG iteration. 
} \label{Fig1}
\end{figure}

The phase diagram shown in \Fig{Fig1} for the extended Bose--Hubbard model is much more complex than for the standard Bose--Hubbard model predicting a normal superfluid (SF) and a Mott insulator phase (MI). The extended Bose--Hubbard model considered here includes both nearest-neighbor interaction $V$ and the interaction-induced tunneling of pairs $P$. For the foremost experimental realization of this model with neutral atoms in optical lattices both amplitudes are equal ($P\:V$) due to the contact interaction potential \cite{Dutta2015}. The effect of the nearest-neighbor interaction $V$ has been extensively studied due to its large amplitude for Coulomb interaction. The pair tunneling is known to cause pair superfluids but the possibility of a ground state with broken time-reversal symmetry induced by interaction has only been recently suggested for a honeycomb lattice \cite{Jurgensen2015}. For one-dimensional lattices, the Hamiltonian of the extended Bose--Hubbard model reads
 \begin{equation}\begin{split}\label{Hamiltonian}
 \hat H = &-J \sum_i \hat a^\dagger_i \hat a_{i+1}  + \mathrm{c.c.} + \frac{U}{2} \sum_i \hat n_i (\hat n_i - 1)  \\
&+2V \sum_{i} \hat n_i \hat n_{i+1} 
+ \frac{P}{2} \sum_i \hat a^{\dagger 2}_i \hat a_{i+1}^2 + \mathrm{c.c.} ,
\end{split}\end{equation}
where $U$ is the on-site interaction energy and $J$ the tunneling amplitude. The bosonic creation and annihilation operators are denoted as $\hat a_i^\dagger$ and $\hat a_i$, respectively, and the occupation of site $i$ corresponds to $\hat n_i = \hat a_i^\dagger \hat a_i$. Let us note that the model neglects next-nearest-neighbor processes and nearest-neighbor density-induced tunneling \cite{Luhmann2012a, Maik2013}, where the first-order effect of the latter can be effectively described as a renormalization of the tunneling amplitude. The correlated pair tunneling $P$ becomes important when the average filling of the lattice is close to $N\:2$ particles per lattice site or higher. As a consequence, the critical $P/U$, where the pair-tunneling-driven twisted superfluid phase emerges, decreases with the filling factor, e.g., $0.07$ for $N\:8$. 

\begin{figure}
\includegraphics[width=\linewidth]{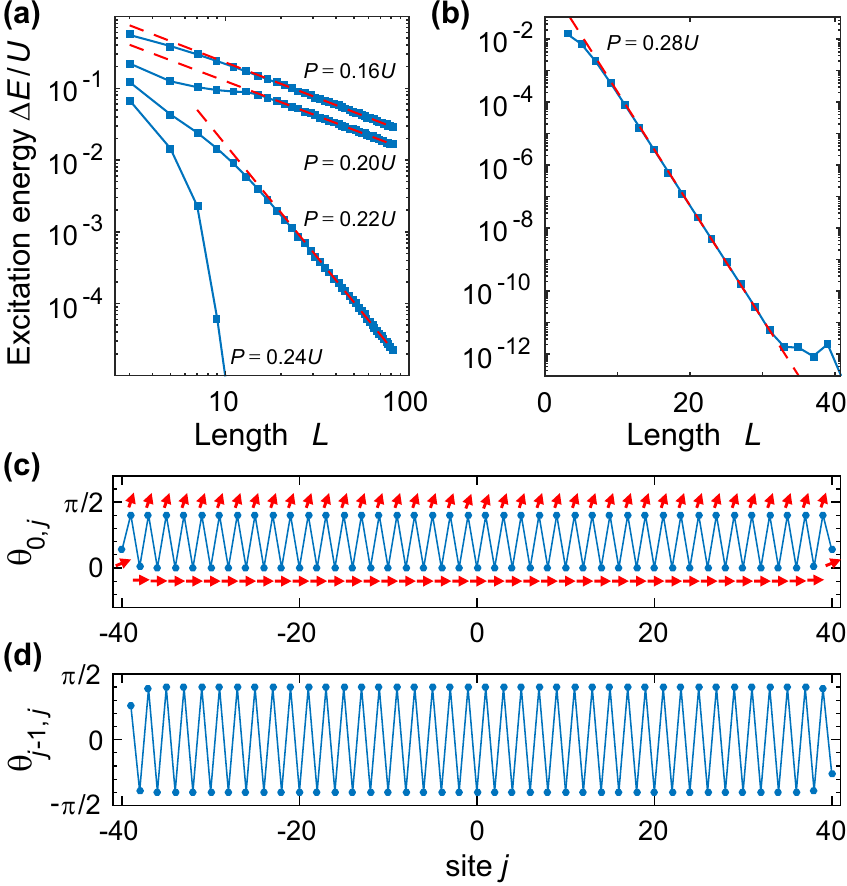}
\caption{Degeneracy of the twisted superfluid state and twisting of the correlation function. 
(a)  Energy gap of the ground state as a function of the system length $L$ within the Mott insulator phase ($N\:4$, $J \: 0.6U$, $P\:0.16U$ to $0.22U$) and in the twisted superfluid ($P\:0.24U$) on a double-logarithmic scale.
(b) For the latter the gap closes exponentially with the length of the lattice.
 (c) In the twisted superfluid phase the complex argument  $\theta_{0,j}$ of the correlation function $g_1(j)$ is alternating for the ground state $\psi_0^\text{tr}$. The red arrows visualize the values of the function $g_1(j)$ in the complex plane. (d) As a consequence, the nearest-neighbor correlation function shows an alternating twisting $|\theta_{j-1,j}|\approx \mathrm{const}$ in the bulk of the lattice with marginal boundary effects. The correlation functions are obtained for $P\: 0.28U$ using the finite-size DMRG algorithm (three full sweeps, 450 density-matrix states).  
} \label{Fig2} 
\end{figure}

The phase boundaries shown in \Fig{Fig1} are computed using the DMRG technique and are depicted for an average filling of $N\:4$ per lattice site. While the superfluid phase, the Mott insulator 
\cite{Kuhner1998,Kuhner2000}, 
the supersolid 
\cite{Sengupta2005,Batrouni2006,Mishra2009,Mazzarella2006,Wessel2007,Gan2007a,Gan2007b,Mathey2009,Maik2013,Jurgensen2015}, 
the pair superfluid phase 
\cite{Kuklov2004,Mathey2009,Hu2009,Menotti2010,Sowinski2012,Jurgensen2015},
and the pair supersolid
\cite{Mathey2009,Hu2009,Zhang2013,Jurgensen2015} 
are expected to appear, the phase diagram shows that a large region in the phase diagram corresponds to the twisted superfluid phase. In this phase the particle number fluctuations are much higher than in the normal superfluid region, which sets a high requirement for the numerical treatment. The DMRG method \cite{White1998,Schollwock2005} is applied for a single center site with 16 Fock states, 450 states for each of the growing system bocks, and two target states for the density matrix (ground state and first excited state). The discarded weight is typically below $10^{-7}$ at the superfluid to Mott transition and reaches up to $3 \times 10^{-5}$ close to the  boundary of the twisted superfluid.  As introduced in \Ref{White2005}, adding corrections to the density matrix is vital for a single center site if using relatively few states to be kept. However, the calculations show that this is not necessary once using a very large number of states (all necessary fluctuations are already present in the environment block). This applies for both the infinite-size algorithm, where the environment block is assumed to be the mirror image of the system block during the growth, and the finite-size algorithm.    

Strictly speaking, the ground state of any bosonic system of finite size is real and non-degenerate according to Feynman's no-node theorem \cite{Feynman1972,Wu2009}. Using an appropriate basis set, the matrix of the extended Bose--Hubbard Hamiltonian \eqref{Hamiltonian} is real and a set of real-valued eigenvectors can be constructed. Consequently, the numerical treatment can be restricted to real numbers and the time-reversal symmetry cannot be broken, since the ground state is nondegenerate. The loophole left in this argument is that the lowest excited state may become quasi-degenerate for a macroscopic system, i.e., degenerate in the thermodynamic limit. An excitation gap that is closing exponentially with the system size indicates therefore a quantum phase with a degenerate ground state. A one-dimensional lattices with $81$ sites is already sufficient for both measuring the energy gap and evaluating long-range correlation functions.

The degeneracy of the ground state in the thermodynamic limit is a requirement for the twisted superfluid phase (TSF). Determining the energy gap of the lowest excitation allows us to distinguish the TSF phase from the surrounding phases, where a threshold value of $10^{-5}U$ is used (filled squares in \Fig{Fig1}). Since the energy gap decreases exponentially across the transition, the phase boundary is only marginally affected by the threshold. The excitation energy as a function of the system size is shown in \Fig{Fig2}a for $J/U\:0.6$ and various values of the pair tunneling $P$ and the nearest-neighbor interaction $V\:P$ across the transition from a Mott insulator to the twisted superfluid. Within the Mott insulator, the excitation energy (with equal particle number) decreases with the system size $L$ approximately as $1/L$, whereas the energy gap vanishes exponentially in the TSF phase (\Fig{Fig2}b). Close to the transition to the TSF phase ($P/U\:0.22$), the gap vanishes with a power-law behavior. In the TSF phase, the energy gap becomes as small as $10^{-12} U$ already for roughly 30 sites, which clearly indicates that the phase transition can be observed in optical lattices with relatively few lattice sites. For larger systems, the numerical error dominates and the energy of the gap fluctuating randomly around $10^{-12} U$ can be considered as degenerate. Higher excitations do not decrease exponentially and are clearly separated from the two lowest states. 

\begin{figure}
\centering \includegraphics[width=\linewidth]{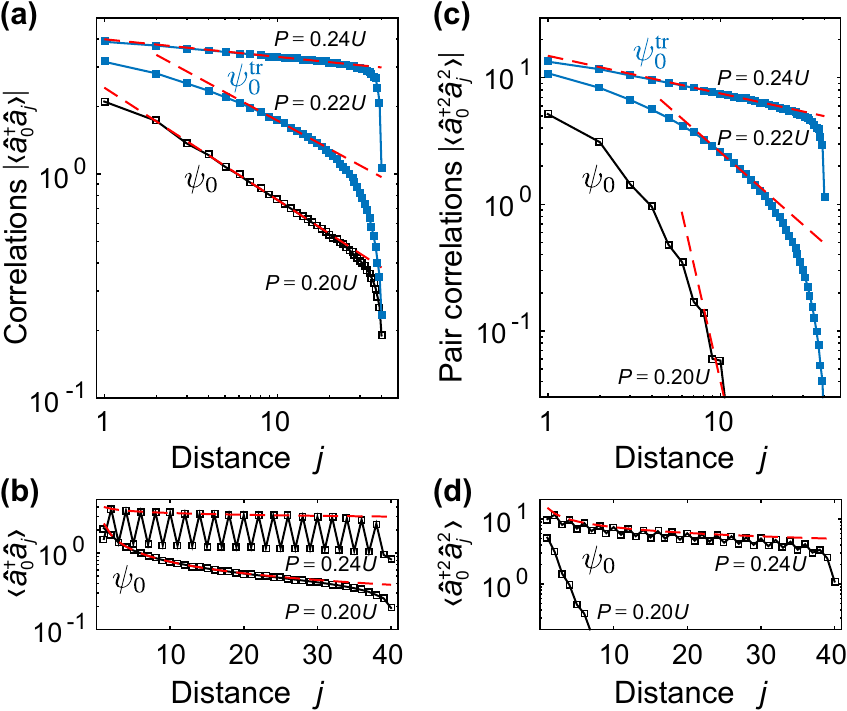}
\caption{ 
Long-range correlation functions across the transition from the Mott insulator ($N\:4$, $J\: 0.6$U, $P\:0.20U$ and $0.22U$) to the twisted superfluid ($P\:0.24U$).
(a) The long-range correlation function $g_1(j)$ on a double-logarithmic and (b) a logarithmic scale. The red lines depict power-law fits $|g_1(j)|\propto (\frac{a}{j})^{\nu_1}$ with $\nu_1 \: 0.5$ ($P \: 0.2U$), $\nu_1 \: 0.43$ ($P \: 0.22U$), and  $\nu_1\: 0.08$ ($P \: 0.24U$). The black lines correspond to the real ground state $\psi_0$ and the blue lines to the complex ground state $\psi_0^\text{tr}$. (c,d)  Double-logarithmic and logarithmic plots of the long-range pair correlation function $g_p(j)$. 
} \label{Fig3}
\end{figure}

From the two real degenerate ground states $\psi_0$ and $\psi_1$ we can build a new pair of ground states that are connected by the time-reversal symmetry operation $\mathrm{T}\psi=\psi^*$. This time-reversal pair is given by $\psi_0^\text{tr}=(\psi_0+i \psi_1)/\sqrt{2}$ and $\psi_1^\text{tr}=(\psi_0-i \psi_1)/\sqrt{2}$. This pair of complex-valued ground states corresponds to the mean-field solution in \Ref{Jurgensen2015} of the twisted superfluid state. In contrast to DMRG, mean-field methods inherently realize the thermodynamic limit and break symmetries, since strongly entangled states such as  $\psi_0$ and $\psi_1$ cannot be described. The nature of the twisted superfluid ground state becomes apparent when evaluating the one-particle correlation function $g_1(j)=\expect{\hat a_0^\dagger \hat a_j}$ of $\psi_0^\text{tr}$ between the central site and a site at a distance $j$. Figure~\ref{Fig2}c shows that the complex argument $\theta_{0,j}= \arg \expect{ \hat a_0^\dagger \hat a_j}$ alternates between neighboring sites, where the red arrows depict the orientation in the complex plane. Between neighboring sites the argument of $e^{i\theta_{0,j}}$ twists back and forth, which is also imprinted on the nearest-nearest neighbor correlations $\theta_{j-1,j}= \arg \expect{ \hat a_{j-1}^\dagger \hat a_j}$ (\Fig{Fig2}d). The twisting is constant on the whole lattice with slight edge effects on the outermost sites. As one would expect, the correlation function $g_1(j)$ of $\psi_1^\text{tr}$  is the complex conjugate of $\psi_1^\text{tr}$ and thus the sign of $\theta_{j-1,j}$ is reversed. Both the twisting and the degenerate ground-state define the twisted superfluid phase.  

Further insight can be obtained by analyzing the correlation functions across the transition from the Mott insulator to the twisted superfluid. Within the Mott insulator, the correlation function $|g_1(j)|\propto (\frac{a}{j})^{\nu_1}$ decreases with an exponent $\nu_1$ larger than $1/4$ (\Fig{Fig3}a). In the twisted superfluid phase, the exponent falls below $1/4$ which is also the case for a normal superfluid. While the correlation function $g_1(j)$ of the real-valued ground state $\psi_0$ varies smoothly in the Mott phase, it alternates strongly close to the phase boundary ($P \: 0.22U$) as well as within the twisted superfluid phase (see \Fig{Fig3}b). Thus, the real-valued ground-state pair $\psi_{0,1}$ shows an amplitude modulation of $g_1(j)$, whereas the unitary transformation to the time-reversal pair $\psi_{0,1}^\text{tr}$ leads to an alternating complex argument and a smooth absolute value (\Fig{Fig3}a). The pair correlation function  $g_p(j)=\expect{\hat a_0^{\dagger 2} \hat a^2_j}$ shows a similar behavior (\Fig{Fig3}c and d), where the complex twisting for $\psi_{0,1}^\text{tr}$ is twice as strong, i.e., $\theta^{p}_{0,j} \approx 2\theta_{0,j}$.    

Several other quantum phases appear in the phase diagram in \Fig{Fig1} which are discussed in the literature for extended Bose--Hubbard models. The phase boundaries between the Mott insulator and the normal superfluid (SF) or the supersolid (SS) are determined by the critical exponent $\nu_1=1/4$ of the long-range correlation function.  The supersolid shows both long-range coherence as well as alternating density correlations $\expect{\hat n_0^\dagger \hat n_j}$
\cite{Sengupta2005,Batrouni2006,Mishra2009,Mazzarella2006,Wessel2007,Gan2007a,Gan2007b,Mathey2009,Maik2013,Jurgensen2015}.
For small tunneling and large pair-tunneling amplitudes, a pair superfluid phase (PSF) appears, where  $\nu_1>1/4$  and $\nu_p<1/4$ indicate a vanishing superfluid order parameter and a nonzero pair order parameter 
\cite{Kuklov2004,Mathey2009,Hu2009,Menotti2010,Sowinski2012,Jurgensen2015}. Due to the positive energy contribution of the pair tunneling $P$, the pair correlation function has a staggered sign $g_p(j)\propto (-1)^j$. Furthermore, a pair supersolid phase (PSS) appears attached to the supersolid phase. The pair supersolid combines pair superfluidity and alternating density correlations \cite{Mathey2009,Hu2009,Zhang2013,Jurgensen2015}. The errors of the phase boundaries are mainly given by systematic uncertainties, which is the fit range for determining $\nu_1$  and $\nu_p$, the finite size of the one-dimensional lattice as well as the threshold value for the energy gap (see Refs.~\cite{Kuhner1998,Kuhner2000} for an error discussion).  

A highly unconventional correlated Mott insulator appears below the pair superfluid phase (\Fig{Fig1}). This phase can be distinguished from the normal Mott insulator by a degenerate ground state, i.e.,  an exponentially closing energy gap (similar as in the TSF phase). Induced by the pair tunneling, the Mott insulator shows staggered pair correlations $g_p(j)\propto (-1)^j$ such as in the pair superfluid phase but with $\nu_p>1/4$. It is therefore denoted here as \textit{pair-wave Mott insulator} (PW).  The breaking of the inversion symmetry resembles the situation of a (charge) density-wave
(e.g. Refs.~\cite{Kuhner1998,Kuhner2000,Sengupta2005,Batrouni2006,Mishra2009,DallaTorre2006,Deng2011,Rossini2012,Mazzarella2006,Wessel2007,Gan2007a,Gan2007b,Mathey2009,Maik2013,Jurgensen2015,Dutta2015,Jurgensen2015}), 
where two possible ground-state occupations exist. In contrast to the staggered $g_p(j)$, the correlation function $g_1(j)$ is positive with $\nu_1\gg 1/4$. In fact, a single-site mean-field treatment would evaluate both the superfluid and the pair superfluid order parameter to zero and hence could not distinguish between this degenerate and a normal Mott insulator phase. 

In conclusion, it is shown that pair-tunneling induces a twisted-superfluid phase in one dimension with an alternating complex phase of the long-range correlation function. The energy gap of the ground state closes exponentially causing quasi-degenerate ground states. The time-reversal  symmetry breaking is caused by interaction-induced pair-tunneling processes,  whereas in chiral superfluids it is induced by degeneracies in the band structure caused either by an effective spin-orbit coupling or by populating excited bands (e.g.  Refs.~\cite{Goldman2014,Wirth2011,Li2014}). The performed density-matrix renormalization group scheme allows a quasi-exact study of a large system, whereas previous studies of the twisted superfluid are either mean-field based \cite{Choudhury2013, Jurgensen2015} or consider only small systems \cite{Cao2014}. While Refs.~\cite{Choudhury2013,Cao2014} found no evidence for the twisted superfluid phase, the results of the cluster mean-field approach in \Ref{Jurgensen2015} is strongly supported by the presented DMRG study.

 As an important contribution to a possible experimental detection of the twisted superfluid phase \cite{Soltan-Panahi2012}, the effect of post-interaction of two different components was discussed in Ref.~\cite{Weinberg2016}. It was found that the interaction within the first milliseconds of the time-of-flight expansion can mimic the signal of a twisted-superfluid in honeycomb lattices. It was shown in \Ref{Jurgensen2015} that for two components the twisted superfluid appears already for much lower values of the pair tunneling due to interaction-induced counter hopping processes and strong correlation effects. Thus, the inclusion of a second spin state in the one-dimensional model is expected to lower the critical value of $P/U$ substantially, where DMRG calculations can shed light on the origin of the underlying correlation effect.

I would like to thank O.~J\"urgensen, C.~Weitenberg, J.~Simonet, and K. Sengstock for various discussions and acknowledge funding by the Deutsche Forschungsgemeinschaft (Grant No. SFB~925).


%

\end{document}